\documentclass[12 pt]{article}
\usepackage{natbib}
\bibpunct{(}{)}{;}{a}{}{,}
\usepackage{graphicx}
\usepackage{txfonts}

\begin{document}
%\unitlength1cm

\title{\LARGE \bf Large discrepancies in error estimates for reverberation times derived from light curves of active galactic nuclei\thanks{Paper dedicated to the memory of Nikolai Gennadievich Bochkarev (19 May 1947 -- 24 December 2022) in appreciation of his encouragement and support of reverberation mapping and his tireless work in the promotion of international collaboration in astronomy.  Nikolai Gennadievich was among the first to recognize the signicance of reverberation mapping and his papers on the topic cover four and a half decades.\\}}
\author{\bf C. Martin Gaskell\thanks{E-mail: mgaskell@ucsc.edu} }
\date{\it  \small  Department of Astronomy and Astrophysics, \\University of California, Santa Cruz, CA 95064}

\maketitle

%%%%%%%%%%%%%%%%%

%\offprints{M.~Gaskell\\ email: \texttt{mgaskell@ucsc.edu}}
%1
%\institute{Department of Astronomy and Astrophysics, University of California, Santa Cruz, CA 95064}

%\authorrunning{Gaskell}
%\titlerunning{Discrepancies in error estimates for AGN reverberation times} %{\it (RN)}}

\date{\small Received 8 November, 2023 November 8 ; accepted accepted 22 April, 2024}

\begin{abstract}
Light-travel-time delays provide one of the most powerful ways of learning about the structure and kinematics of active galactic nuclei (AGNs).  Estimating delays from observations of AGN variability presents statistical challenges because the time series are almost invariably irregularly sampled. Correct assessment of errors in lag estimates is important for evaluating results. The most widely used method of determining phase lags has been the interpolated cross-correlation function method introduced by Gaskell and Sparke (1986).  It is shown here that the widely used modified smooth bootstrap method of Peterson et al.\@ (1998) significantly overestimates the error in lags derived using the Gaskell \& Sparke (GS) method, especially for poorly sampled light curves. The remarkably high accuracy claimed for lags obtained by the JAVELIN method of Zu et al.\@ (2011) (more than an order of magnitude improvement for 25\% of cases) is spurious and the accuracy no better than the GS method. A related method proposed by Li et al.\@ (2013) suffers from similar but less serious problems.  A slightly modified version of the analytic formula of Gaskell and Peterson (1987) gives an easy and accurate way of estimating lag errors for well-sampled, high-quality data.  The width of the continuum autocorrelation function is shown to be proportional to the square root of the luminosity. This is useful in planning observing campaigns.  The discrete correlation function method is less powerful than the GS method and for many typical situations gives errors in the lag twice as large as those of the GS method.
\\

\end{abstract}

\maketitle

\section{Introduction}\label{sec1}

The discovery of variability of the broad-line emission in an active galactic nucleus (AGN) by \citet{Dibai+Esipov65} and \citet{Burbidge+Burbidge66} (see also \citealt{Wampler67}), prompted \citet{Shklovskii66} to recommend the undertaking of prolonged, systematic observations of broad-line variability as a means of learning about AGNs.  Following this suggestion, \citet{lyutyi+cherepashchuk72} and \citet{cherepashchuk+lyutyi73} showed that the size of broad-line regions (BLRs) in active galactic nuclei (AGNs) could be estimated from the phase lag between continuum variability and line variability because the emission of reprocessed radiation is delayed by the extra time it takes the direct, unreprocessed, driving radiation to reach the reprocessing regions. The first study of the energetics of the broad-line variability was by \citet{Bochkarev+Pudenko75} who showed that the necessary extreme ultraviolet (EUV) to soft X-ray energies needed for photoionizing and heating the gas were consistent with observations of the high-energy spectrum.

In an idealised case, the light curve of the reprocessed radiation is the light curve of the driving radiation convolved with an impulse response function, $\Psi(\tau)$, where $\tau$ is the time delay (see \citealt{Blandford+McKee82} for details).  The impulse response function (IRF) is the response of a system to a $\delta$-function\footnote{Following \citet{Blandford+McKee82}, $\Psi(\tau)$ has been commonly called the ``transfer function" in AGN literature, but, as \citet{Uttley+14} point out (see their footnote 1), in signal processing the  term ''transfer function" is used for the Fourier transform of the IRF.}.  Gaining information about AGNs from analysis of the lags in light curves is now referred to as ``reverberation mapping''.  The easiest quantity to estimate, and one of the most important, is the phase lag, the first moment of $\Psi(\tau)$.  Even determining this is not straight forward, however, since AGNs time series are rarely uniformly sampled.  The most widely used method of estimating the lag from AGN light curves has been the partial-interpolation cross-correlation function technique introduced by \citet{Gaskell+Sparke86}.  This method deals with the problem of irregular sampling by cross correlating the observed points in one time series with observed or linearly interpolated points in the other time series (see Fig.~1 of \citealt{Gaskell+Peterson87}). The lag is taken to be either the peak of the cross-correlation function (CCF) or the centroid of the peak above some threshold (usually taken to be 80\% of the peak of the CCF). Gaskell \& Sparke (GS) showed that by using this technique BLR sizes could be obtained in a single observing season for a typical Seyfert galaxy. The GS method has therefore been used extensively for measuring the sizes of BLRs.  

The method has also been used to show that the BLR is gravitationally bound. \citet{Bochkarev+Antokhin82} and \citet{Antokhin+Bochkarev83} pointed out that broad-line profile changes on a timescale of days should be observable and observations of them would distinguish between different kinematic models for the BLR. In 1983 Bochkarev initiated what was to become a very successful long-term AGN monitoring campaign with the 6-m telescope of the Special Astrophysical Observatory to study broad-line profile changes. This has has produced over 40 journal papers published to date.  The monitoring quickly showed that there were indeed rapid changes in broad-line profiles \citealt{Bochkarev84}.  
\citet{Gaskell88} introduced velocity-resolved reverberation mapping to determine the direction of gas motion by comparing the variability of red and blue wings of lines (see also \citealt{Koratkar+Gaskell89,Koratkar+Gaskell91}).  This gave the important result that the BLR gas is virialized and thus opened up the estimation of black hole masses from reverberation mapping (see \citealt{Marziani+Sulentic12} for a review of mass estimation from the BLR and \citealt{Kormendy+Ho13} for a general review of supermassive black hole masses).  There have been many other applications of the GS technique.  It is used to measure the radius of the hot inner edge of the dusty torus \citep{Clavel+89} and of a scattering region producing polarized flux \citep{Gaskell+12}.  Lags have also been determined for the extreme UV and X-ray regions (e.g., \citealt{Chiang+00,Zoghbi+11,Zoghbi+12}). Much of our knowledge of AGNs and the masses of supermassive black holes thus rests on the results of reverberation mapping.  Furthermore, \citet{Oknyanskij02} has suggested that the reverberation mapping could be used to estimate cosmological parameters by giving the distances to AGNs.

A result of \citet{Gaskell+Sparke86} that was surprising at the time was to confirm that the BLR was smaller than had been expected from pre-1986 photoionization models.  This naturally led to questions being asked about the errors in the determination of the lags.  In particular \citet{Edelson+Krolik88} argued that the approach of \citet{Gaskell+Sparke86} did not give a meaningful size of the H$\beta$-emitting region of Akn~120 and that the size of the region was consistent with their photoionization prediction of 200 days (for $H_0$ = 70 km s$^{-1}$ Mpc$^{-1}$) rather than the $14 \pm 12$ days \citet{Gaskell+Peterson87} estimated.  This would require errors in lags determined by \citet{Gaskell+Peterson87} to be underestimated by more than an order of magnitude.  The focus of studies of the sizes of the errors in the late 1980s and early 1990s was thus on whether the errors in the GS method were indeed more than an order of magnitude larger than thought.

\citet{Gaskell+Peterson87} gave a simple analytic formula (their Eq.~4) for estimating errors. This will be discussed below (see section 10).  To investigate errors further they introduced the modelling of AGN light curves as damped random walks in order to produce artificial light curves for estimating errors via Monte Carlo simulations.  Subsequent work \citep{Kelly+09,Kozlowski+10,MacLeod+10,Kelly+14} confirms that a damped random walk is indeed a good approximation to AGN variability. The simulations of \citet{Gaskell+Peterson87} showed that the errors were non-Gaussian in the sense that the probability of a large error was larger than would be expected from a normal distribution.  This was also found by \citet{Maoz+Netzer89} who used resampling of smoothed observed light curves to estimate errors.  An example of the extended non-Gaussian tail of the distribution can be seen in Fig.~4 of \citet{White+Peterson94} where it can also be seen that the tail becomes more prominent as the sampling gets worse.  Because of this, the analytic formula of \citet{Gaskell+Peterson87} will underestimate the error in the lag when observational errors are large or the sampling is poor.  The Monte Carlo simulations of \citet{Maoz+Netzer89} and \citet{Koratkar+Gaskell91} showed that the Gaskell \& Peterson analytic formula can indeed seriously underestimates the error when the sampling is poor.

A model-independent indication of the errors in lag determinations was given by simply comparing results from independent subsets of the data \citep{Gaskell88}.  This simple, rough, independent check gave support to the idea that the other error estimates were not systematically underestimating the true errors by over an order of magnitude \citep{Gaskell88,Koratkar+Gaskell89,Koratkar+Gaskell91}.  The Monte Carlo simulations of \citet{Gaskell+Peterson87} were extended by \citet{White+Peterson94} to explore the effects of different continuum power density spectra, different shape response functions, and the coarseness of the sampling.  All of the above mentioned studies ruled out order-of-magnitude errors in the determination of the errors in the lags.  Since the 1980s, the quantity and quality of reverberation mapping data have improved significantly and the re-observations have generally confirmed earlier results.  For example, for Akn~120 \citet{Peterson+Gaskell91} found an H$\beta$ lag of $39 \pm 14$ days for the lag of H$\beta$ in Akn~120 and with later data \citet{Peterson+04} give $47.1 \pm 10.3$ and $37.1 \pm 5.1$ days for two different epochs.  All these are $\ll 200$ days, the size given by \citet{Edelson+Krolik88} from photoionization modelling.

Although it was agreed early on \citep{Gaskell+Peterson87} that errors in lags were best determined via Monte Carlo simulations, such error estimates are not completely model independent as it is necessary to assume a form of the continuum variability and the shape of $\Psi(\tau)$.  To avoid the possible effects of these assumptions \citet{Peterson+98} refined the method of estimating errors from subsets of the data.  This has the advantage that the error estimates are based solely on the data themselves.  Because of its ease of use, it is has been the method most widely used to calculate errors.

Although there is no longer any controversy over the sizes of BLRs being substantially smaller than expected, the question of the accuracy of the estimation of errors in lags remains of interest.  An accurate knowledge of the errors is important in knowing how much scatter in quantities and relationships derived from reverberation mapping is due to observational errors and how much is intrinsic.  An accurate knowledge of errors is also important for good stewardship of observing time when planning observing campaigns.  It is also highly desirable, of course, to use analysis methods which minimize errors in the lag.

Interpolating in the data space as in the GS method is only one of many methods that have been considered for determining lags.  \citet{Clavel+87} suggested simply visually comparing light curves.  More quantitatively, \citet{Van_Langevelde+90} proposed that the lag be determined by shifting the phase between two time series until one obtains the smoothest combined time series.  Shifting light curves can be done most easily by measuring the lag between corresponding peaks or troughs in the two light curves \citep{Litchfield+95}.  \citet{Press+92} discuss obtaining the lag from the smoothest combined time series by global $\chi^2$ minimization and \citet{Rybicki+Kleyna94} extended this to estimating higher moments of $\Psi(\tau)$.  Another approach is parametric modeling of the time series using maximum likelihood fitting of a damped random walk to the non-uniformly sampled data \citep{Jones81}.

More recently \citet{Zu+11} have developed a variant of these methods where they assume a form of $\Psi(\tau)$ and estimate the errors in the lag by a Markov chain Monte Carlo (MCMC) approach.  They initially called this method SPEAR but the name was changed to JAVELIN.  A somewhat similar approach has been used by \citet{Li+13} who, rather than assuming $\Psi(\tau)$, assume a physical model of the BLR.  They also estimate errors using a MCMC analysis. As will be shown below, the published errors bars of lags obtained from JAVELIN are on average {\em substantially} smaller than those given by the GS method.  If the error estimates are correct, then the JAVELIN method represents a significant advance in reverberation mapping.  The errors in the lags given by \citet{Li+13} are also significantly smaller than those given by the GS method, although not by as large a factor as the error estimates from JAVELIN.

Motivated by the apparent major improvements offered by the JAVELIN method and the method of \citet{Li+13}, this paper first re-examines the accuracy of estimates of errors for the \citet{Gaskell+Sparke86} method (which has been the most widely used method for determining lags) and then investigates whether the methods of \citet{Zu+11} and \citet{Li+13} do indeed give substantially smaller errors than the traditional GS method.  The simple analytic formula given by \citet{Gaskell+Peterson87} is examined.  Obviously, an important reason for understanding errors in lags is to be able to plan observing campaigns.  A prescription is given for roughly estimating the expected accuracy in advance.  Other methods of estimating lags are also mentioned and, in particular, the shortcomings of the discrete correlation function method are discussed.

It should be noted that the purpose of this paper is only to examine the accuracy of lags determined by methods that have been applied to a large number of AGNs, not to review or evaluate all methods used in the analysis of unevenly sampled time series, or such topics as the search for periodicities, optimal determination of the shape of auto-correlation or cross-correlation functions, or the recovery of $\Psi (\tau)$.  Some reviews of methods of treating irregularly sampled data can be found in \citet{Babu+Stoica10} and \citet{Rehfeld+11}.

\section{Using subsets of data to estimate errors}

\citet{Gaskell88} introduced the use of independent halves of the time series to get a model-independent assessment of the errors.  The most extensive compilation of errors estimated by this method is in \citet{Koratkar+Gaskell91}.  This method was only intended as a rough check that the analytic formula of  \citet{Gaskell+Peterson87} and Monte Carlo simulations were not underestimating the errors in the lags by an order-of-magnitude. \citet{Peterson+98} improved on the use of subsets of the data with the introduction of a modified smooth bootstrap method.\footnote{Smooth bootstrapping refers to adding noise to the data during the bootstrap procedure.  \citet{Peterson+98} refer to their method as FR/RSS which stands for ``flux randomization with random subsets''.}  In bootstrapping a set of $n$ observations is sampled $n$ times and the random sample is used to calculate the quantity of interest.  Each random sample will omit some observations but sample others more than once.  The problem with doing this for pairs of observations from two time series is that the time of the observation is critically important in time-series analysis and counting the pair of observations on one night more than once is redundant.  \citet{Peterson+98} solved this problem by counting each duplicate as one observation. Because of this the effective sample size is reduced by $\approx 37$\% \citep{Peterson+98} so the total number of observations used on average is only 63\% of the original number.  It is possible to increase the weighting for a night chosen more than once (see \citealt{Welsh99}) but this only reduces the size of the estimated errors by $\approx 3 - 6$\% (see Appendix A of \citealt{Peterson+04}). The problem of the reduced number of epochs remains.  For the rough method of \citet{Gaskell88} the reduction in the number of observations used is 50\%.  The effects of this were ignored by \citet{Gaskell88} and subsequent papers because the aim was to give an independent check on the results of the Monte Carlo simulations and conservative independent estimates were useful in establishing that the errors were not being grossly underestimated.  \citet{Peterson+98} also ignored the systematic effect of the reduction of the sample size.  They noted that their smooth bootstrap error estimates were systematically greater than Monte Carlo estimates, but they considered more conservative errors estimates to be desirable.

Under ideal circumstances, one might expect the precision of the determination of a quantity to improve as $1 / \surd n$ as the number of observations, $n$.  For the 50\% loss of sample size in the rough method of independent samples of \citet{Gaskell88} this would result in errors being overestimated by a factor of $\surd 2$. This was considered unimportant given that, as noted, the method was merely a rough independent verification that the errors were not being underestimated by an order of magnitude, as would be needed to reconcile the Ark~120 BLR size estimates with the 200 light-day size \citet{Edelson+Krolik88} proposed from photionization considerations.  However, as we will see, the decline in accuracy is worse than $1 / \surd n$ for small $n$.

The effect of the reduced number of epochs can studied via Monte Carlo simulations using artificial light curves.  \citet{White+Peterson94} studied the effect of reducing the sampling by more than a factor of six.  Fig.~1, based on their simulations, shows the increase in the logarithm of the error, $e$, versus $\log n^{1/2}$.  One can see from this figure that for well-sampled light curves (large $n$) the change in the error is consistent with simply being proportional to $n^{-1/2}$ (the solid line in Fig.~1).  However, it is also apparent that, for ill-sampled light curves (the left side of Fig.~1), the increase in the error is faster than $n^{-1/2}$ as $n$ decreases.  From these simulations the increase in $e$ for small $n$ is consistent with $e \propto n^{-1.3}$.

% Figure 1
 \begin{figure}
   \centering \includegraphics[width=10cm]{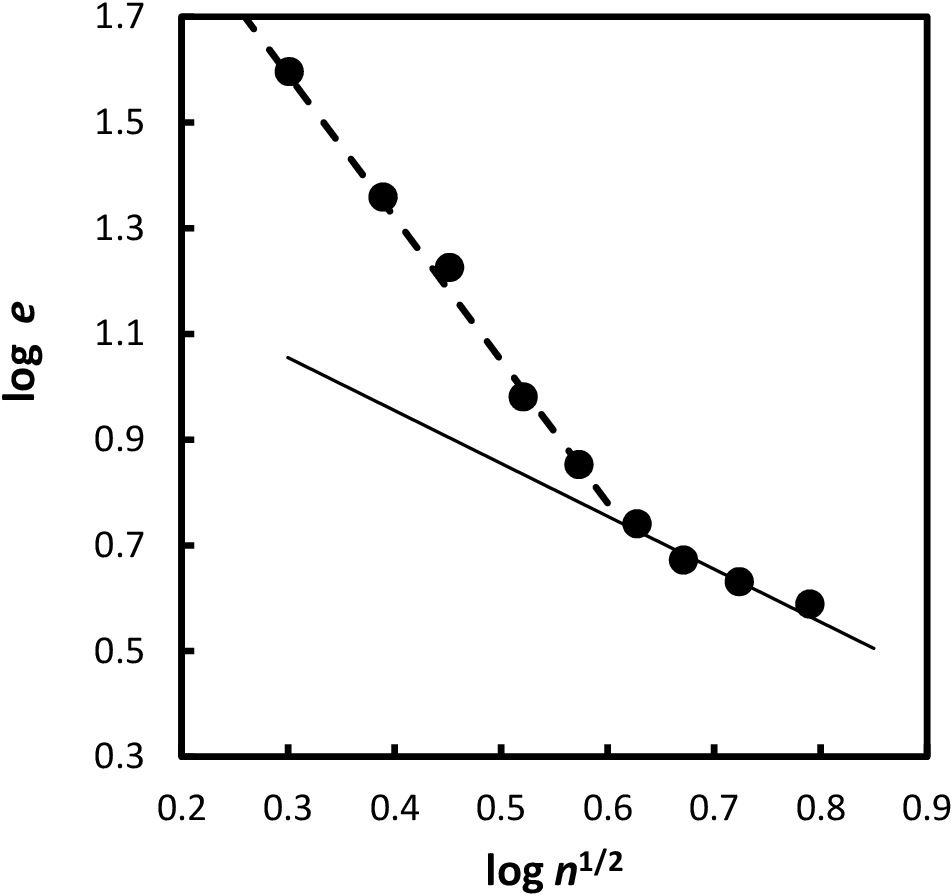}
      \caption{The error, $e$, in lag estimates as a function of $\log n^{1/2}$, where $n$ is the number of observations. The points are the averages of the four 20-light-day BLR simulations in Tables 1 and 2 of \citet{White+Peterson94}.  The solid line, fit to the four simulations with the largest number of observations, shows $e \propto n^{-1/2}$.  It can be seen that for well-sampled light curves the increase in the error is consistent with this.  The dashed line is an $e \propto n^{-1.3}$ fit to the simulations with a small number of observations. It is apparent that for ill-sampled light curves (small $n$) the rate of increase in the errors with decreasing $n$ is faster than $n^{-1/2}$.}.	
       \label{Fig1}
  \end{figure}
% The data for this figure is in RSSFR_testing - White+Peterson94

\citet{Peterson+98} give the results of Monte Carlo simulations and the corresponding smooth bootstrap error estimates for increasingly poor sampling. Table 1 gives the mean ratios of the errors calculated via their smooth bootstrap procedure (FR/RSS) to the errors derived from Monte Carlo simulations as a function of, $n$ (data from Table 4 of \citealt{Peterson+98}).  The third column gives the expected ratio of the two error estimates based on Figure 1.  $n = 40$ observations is on the right-hand side of Fig.~1, so the increase in error is $0.63^{-0.5}$ and $n = 10$ is on the left, so the increase in the error is $0.63^{-1.3}$.  For the intermediate case the dependence has been taken as the geometric mean of the two extremes.   The increase in the ratio of the error estimates is in agreement with the predictions.

\begin{table}
\centering
\caption[]{Ratio of errors calculated via the smooth bootstrap procedure of \citet{Peterson+98} to errors derived from Monte Carlo simulations as a function of number of observations.\\}
\begin{tabular}{ccc}
\hline \hline
 $n$ &    FRRSS/MC    & Predicted \\
\hline
  40 & $1.34 \pm 0.11$ & 1.36     \\
  20 & $1.45 \pm 0.09$ & 1.45     \\
  10 & $1.94 \pm 0.08$ & 1.82     \\
\hline
\end{tabular}    \\
\label{tab1}
\end{table}

By the same reasoning, error estimates from independent halves of the data are expected to be worse by at least $1 /\surd 0.5 = 1.40$.  For the 33 error estimates tabulated by \citet{Koratkar+Gaskell91} the geometric mean ratio of errors from the subsets to errors from Monte Carlo simulations is 1.47.  These are thus in good agreement with the predicted overestimate.

\section{Peak or Centroid?}

With the GS method it has been common practice to quote both the lag of the peak of the cross-correlation function and the centroid.  For a physically reasonable  asymmetric response function the centroid lag will be greater than the peak lag, but in practice the difference is small.  Inspection of the large compilation of lags of \citet{Peterson+04} shows that while there is a significant ($p \approx 0.02$) difference in the sense expected, the effect is small ($\thickapprox 6$\%) so that it is indistinguishable in an individual object and can be ignored in practice.  However, because the location of the peak of a cross-correlation function is more sensitive to noise in the cross-correlation function than the centroid is, it is expected that the error in the peak, $e_{Peak}$, will be greater than the error in the centroid, $e_{Cent}$.  This is supported by Monte Carlos simulations (see Fig.~1 of \citealt{White+Peterson94}). Inspection of the Monte Carlo simulations with artificial light curves \citet{Peterson+98} shows that the geometric mean of $e_{Peak} / e_{Cent} \approx 1.3$.  For real AGN observations, inspection of the large compilation of lag estimates of \citet{Peterson+04} calculated by their smooth bootstrap method gives the geometric mean of $e_{Peak} / e_{Cent} \approx 1.2$.  Both simulations with artificial data and a large number of actual observations therefore confirm that the lag of the peak is about 25\% less accurate that the lag from the centroid.   There is thus no need to give peak lags.

\section{JAVELIN}

One well-known source of uncertainty in reverberation mapping is what the continuum is doing when the AGN is not being observed.  In the \citet{Gaskell+Sparke86} approach the continuum in a gap is taken to be a linear interpolation between the observed points on either side of the gap and deviations from this are just taken to be an additional random error (see Figure 1 of \citealt{Gaskell+Peterson87}).  The accuracy of this depends on the smoothness of the light curve.  Two important points are frequently overlooked.  The first is that in the GS method the interpolation only needs to be done in {\em one} time series.  The second is that for the BLR the line time series is the continuum time series convolved with $\Psi$($\tau$) and therefore the {\em line} time series will be {\em smooth} on the timescale of the light-crossing time for that line \citep{Gaskell+89}.  This is because a convolution is a smoothing (blurring) when $\Psi$($\tau$) $\geq 0$ for all $\tau$.  The resulting smoothness of the line light curve can be appreciated by looking at Figure 3 of \citet{White+Peterson94}.  In practice, the continuum light curve is already relatively smooth as well (e.g., the optical light curved is smooth on the light-crossing time of the H$\beta$-emitting region).  This is because the power density spectrum, $P(f)$, of the optical/UV variability of AGNs is very red: $P(f) \propto 1 / f^{\alpha}$ where $f$ is the frequency of the variability and $\alpha \sim 2$ (see, for example, \citealt{White+Peterson94}\footnote{\citet{Edelson+Krolik88} assumed, in the absence of data, that AGN variability was flicker noise ($\alpha \sim 1$) and thus their simulations have much larger short-period variability than is actually the case.}).  For a damped random walk assumed by \citet{Gaskell+Peterson87}, $\alpha \sim 2$.

Rather than using the linear interpolation of \citet{Gaskell+Sparke86}, \citet{Zu+11} modelled the gaps in the observations using many light curves produced from damped random walks.  It should be noted that this is {\em not} the same as the use of a damped random walk by \citet{Gaskell+Peterson87}, \citet{Koratkar+Gaskell91}, etc.  The latter used the damped random walk to generate complete artificial continuum light curves which were sampled with the same sampling as the real data, while \citet{Zu+11} use the damped random walks only to fill in the gaps in the data and to estimate the error bars (see Figure 2 of their paper for an illustration).  This enables \citet{Zu+11} to do a detailed analysis of the errors due to the interpolation process. To calculate the phase lags, \citet{Zu+11} convolved their reconstructed light curves with an assumed response function and compared the results with the observed line light curve.  The errors in the lag estimates are estimated via a MCMC (Markov chain Monte Carlo -- see \citealt{Zu+11} for details).

\begin{figure}
   \centering \includegraphics[width=10cm]{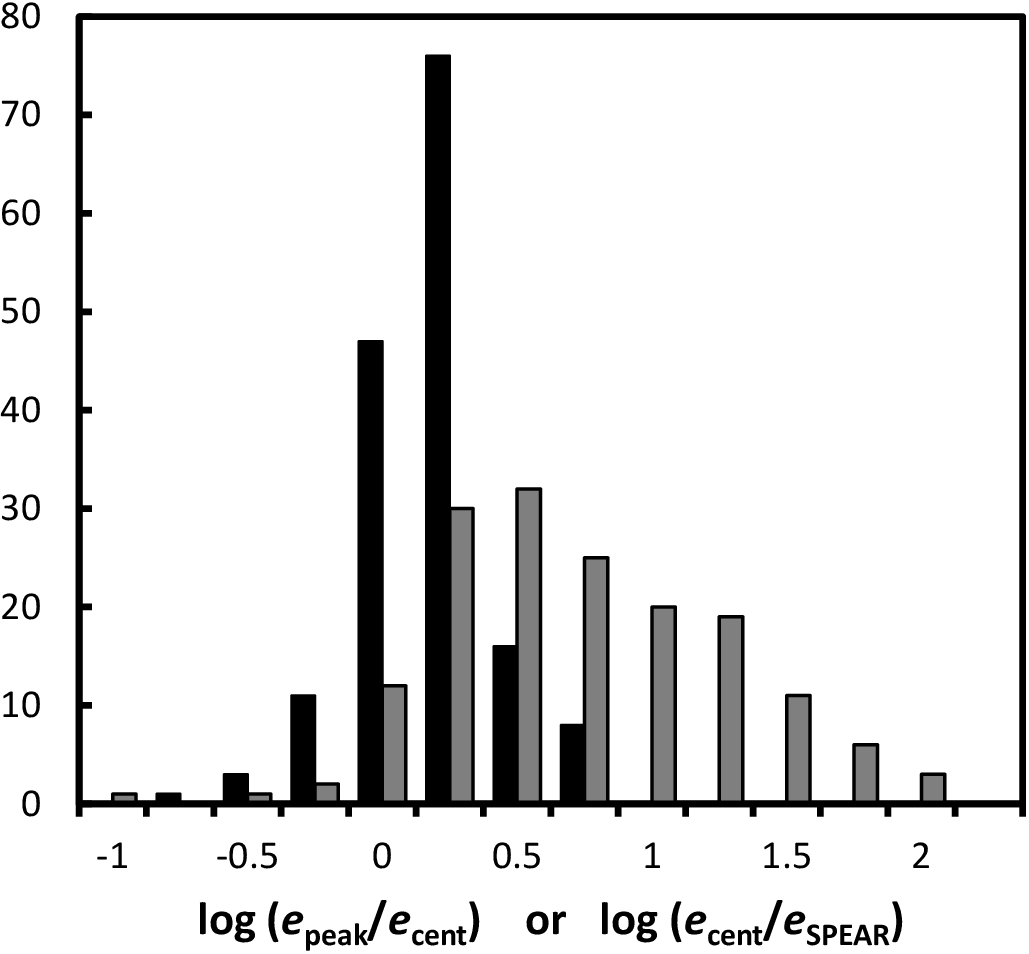}
      \caption{Histograms of the ratios of the distributions of the logarithms of the ratios of different error estimates.  The binning is in units of 0.25 dex and the labels refer to the upper limit of each bin.  The black bars show the distribution of the logarithms of the ratios of the errors in the peaks of the cross-correlation function given by the GS method, $e_{peak}$, to the corresponding errors in the centroids, $e_{cent}$.  It can be seen that the black historgram is not quite centered on 0 and this implies that the $e_{peak}$ is systematically larger than $e_{cent}$ (as discussed in section 3).  The grey bars show the distribution of the logarithms of the ratios of $e_{cent}$ to the corresponding errors, $e_{JAVELIN}$, given by JAVELIN.  It can be seen that for a large fraction of cases the errors given by the JAVELIN method are more than an order of magnitude smaller than those given for the GS method.}.	
       \label{Fig2}
  \end{figure}

\citet{Zu+11} give 87 lag estimates for most AGNs that had been well observed at the time.  These include different lines and multiple epochs.   There are 174 error estimates.  \citep{Peterson+04} have produced homogenized lag estimates using the GS method for almost all of the lines.  They give lags from both the peak ($\tau_{peak}$) and the centroid ($\tau_{cent}$) of the cross-correlation function and estimates of the errors, $e_{peak}$ and $e_{cent}$, in the respective lags using a slightly modified version of the smooth bootstrap method of \citet{Peterson+98}.  The black bars in Fig.~\ref{Fig2} show the distribution of the logarithms of ratios of $e_{peak}$ and the corresponding $e_{cent}$.   It can be seen that the distribution is not quite symmetric about $\log e_{peak}/e_{cent} = 0$.  This is because, as discussed in section 3, $e_{peak}$ is slightly greater than $e_{cent}$.

% Figure 2
% This figure is from

The grey histogram in Fig.~\ref{Fig2} shows the distribution of the logarithms of ratios of $e_{cent}$ to the corresponding error estimates, $e_{JAVELIN}$, obtained by \citet{Zu+11} from JAVELIN.  It can be seen that, except for a few cases, $e_{JAVELIN}$ is much smaller than $e_{cent}$.  The geometric mean of $e_{cent} / e_{JAVELIN} = 4.2$.  Furthermore, as can be seen in Fig.~\ref{Fig2}, the distribution of $e_{cent} / e_{JAVELIN}$ is highly asymmetric even in a logarithmic plot.  In fact, the upper quartile of the distributions of $e_{cent} / e_{JAVELIN}$ is 10.  Thus, {\em if the error estimates of \citet{Zu+11} are taken at face value, they imply that 25\% of the time JAVELIN does more than an order of magnitude better than the GS method.}  To take an example, for the broad H$\beta$ line of Mrk 509 optical monitoring has an average cadence of 15-18 days or a median cadence of 7 days -- quite typical for a reverberation-mapped AGN.  \citet{Zu+11} give this their top quality rating (their group A).  The GS method gives a lag of $79.6 \pm 5.7$ days (centroid) or $76 \pm 6.5$ days (peak) with the errors determined by the smooth bootstrap method of \citet{Peterson+98}.  JAVELIN gives a lag for H$\beta$ in agreement with this (69.9 days), {\em but with an error bar of only $\pm 0.3$ days (i.e., about $\pm$8 hours).}

\section{Are the errors given by the JAVELIN method accurate?}

Since the observation sets analyzed by \citet{Zu+11} also have peak and centroid lags given by \citet{Peterson+04} using the GS method with errors determined by the smooth bootstrap method, and since the errors in the GS method are well understood (see section 2), we can compare the agreement between the GS method and JAVELIN.  If the estimates are independent and the error estimates for each method are correct, the differences between estimates for the same lines in the same objects should give a $\chi^2 / dof \approx 1$. If we first look at the differences between the peak and centroid lags given by the GS method we find $\chi^2 / dof = 0.2$ using the quoted errors in the peaks and the centroids.  If we reduce $e_{peak}$ and $e_{cent}$ by a factor of 1.36 (see Section 2) we find $\chi^2 / dof = 0.3$ which is still $\ll 1$.  To get $\chi^2 / dof = 1$ we would need to reduce the errors in $e_{peak}$ and $e_{cent}$ by a factor of 2.3 and the errors in lags from the GS method are not overestimated by such a large factor.  The real reason $\chi^2 / dof \ll 1$ is, of course, that the peak and centroid lags are {\em not} independent; they are from the same data set.

If we perform a similar analysis comparing the centroid from the GS method with the JAVELIN method we get $\chi^2 / dof = 11.8$.  If we assume that the error estimates given by JAVELIN ($e_{JAVELIN}$) are correct, the errors in the centroid would need to be increased by a factor of 4.2 to get $\chi^2 / dof = 1$ (or by an even larger factor if the lags determined by the GS method and by JAVELIN are not independent).  Given the evidence above that the \citet{Peterson+98} modified smooth bootstrap method {\em over}estimates $e_{cent}$ by a factor of 1.36 or more, such a large systematic underestimate of $e_{cent}$ is most unlikely.  If instead we assume that $e_{cent}$ needs to be reduced by a factor of 1.36, then it is $e_{JAVELIN}$ which is underestimated and it needs to be increased by a factor of 9.6.  This comparison thus argues that rather than JAVELIN being, on average, 4.2 times better than the GS method, {\em the real uncertainties in the lags JAVELIN gives are about twice as large as for the GS method.}

\section{External checks on errors}

The comparisons discussed in section 5 are only between errors calculated by different methods applied to the same data. To learn which error estimate is most accurate we need an external check -- i.e., some {\it a priori} knowledge of what the lags ought to be.

\subsection{The lag predicted from luminosity}

One possibility is to look at the relationship between the effective radius of the region of the BLR emitting a given line (e.g., H$\beta$) and the line or continuum luminosity which gives an independent prediction of the size of the BLR \citep{Dibai77}.  \citet{Zu+11} show a comparison of the radius-luminosity relationship for AGNs with H$\beta$ radii estimated by the GS method and by JAVELIN (see their Fig.~9).  If there is no intrinsic scatter in the radius-luminosity relationship, a tighter relationship when the radii are estimated using JAVELIN would provide good support for the superiority of JAVELIN.  However, the scatter about the relationship is the same ($\approx 0.21$ dex) for two methods (see Table 2 of \citealt{Zu+11}). The $\chi^2$/~$dof$ for the radius-luminosity relationship using the JAVELIN lags is much greater than the $\chi^2$/~$dof$ for the relationship using lags determined by the GS method but this is simply because the error bars given by JAVELIN are smaller.  The $\chi^2$/~$dof$ values would be similar if the errors in $e_{JAVELIN}$ were larger.  Beyond this it is not possible to come to a strong conclusion about accuracy of the error bars given by the two methods because of the probability (which \citealt{Zu+11} note) of intrinsic scatter in the radius-luminosity relationship because of differences between objects.  One difference that has to be taken into account is reddening, \citet{Zu+11} use luminosities that are not corrected for internal reddening in the AGNs.  It is now clear that internal reddening of AGNs is substantial (see \citealt{Gaskell17} and \citealt{Gaskell+23})  A further problem is that, for a given object, the lags vary with time.  This is very common and is discussed at length in \citet{Gaskell+21}.

\subsection{Repeated observations of the same object}

The problem of intrinsic object-to-object scatter in the radius-luminosity relationship because of object-to-object differences could be circumvented in principle with repeated observations of the same object at the same average flux level so long as the geometry does not change with time.  However, when an AGN is re-observed it is unlikely to be at the same flux level it was at when previously observed and the size of the BLR is expected to vary with the mean flux level (this change in size is sometimes referred to as ``breathing'').  The AGN which has been most studied by far with reverberation mapping is NGC~5548.  The BLR radius of NGC~5548 varies as approximately $L_{opt}^{0.5}$ \citep{Peterson+02,Cackett+Horne06}.  This is what is expected in the self-shielding BLR model of \citet{Gaskell+07} if the overall continuum shape is the same for high and low luminosities.\footnote{Note that such a dependence is {\em not} expected for cloud models such as the locally optimized cloud (LOC) model of \citet{Baldwin+95} because in the LOC model only the Str\"omgren lengths in the individual clouds change and these changes are very small compared with the distance of each cloud from the center of the system.  For a comparison of the LOC model and the self-shielding model, see \citet{Gaskell09}}  This luminosity dependence needs to be allowed for in comparing lags obtained at different epochs.  For the high-quality, repeated observations of NGC~5548 we not only have lags determined by the GS method and JAVELIN, but also by fitting $\Psi(\tau)$ \citep{Cackett+Horne06}. The median lag from the response function ($\tau_{med}$ of \citealt{Cackett+Horne06}) can be directly compared with the lag given by the centroid of the CCF.  The average error of $\tau_{med}$ is $1.6 \pm 0.1$ times smaller than the average error of the lag from the centroid of the CCF when the errors are estimated by the modified smooth bootstrap method of \citet{Peterson+98}, but given that the latter is overestimating errors by at least a factor of 1.36 (see section 2), the advantage of the \citet{Cackett+Horne06} approach is at best $18 \pm 7$\%.  Comparison of the four estimates of the NGC~5548 lags for each of season with the radii predicted by the corresponding mean optical continuum luminosity or H$\beta$ luminosity (i.e., allowing for the ``breathing") gives a scatter of $\sim 0.11$ dex. This is almost half the scatter in the radius-luminosity relationship for different reverberation-mapped AGNs (Figure 9 of \citealt{Zu+11}), but the scatter does not vary significantly for the four different sets of lag estimates (probability $p > 0.4$ for the differences arising by chance).  The insensitivity of the scatter to the method used to determine the lags again suggests that the scatter is dominated by intrinsic scatter in NGC~5548 rather than errors in estimating the lags. As \citet{Gaskell+21} have show for NGC~5548 and many other reverberation-mapped AGNs, significant differences in lags can be seen for consecutive flaring events in the same object.  Changes on timescale of a typical continuum event will be due to the off-axis nature of the variability of AGNs \citep{Gaskell09,Gaskell10,Gaskell11}.  Changes on longer timescales can also arise from varying partial obscuration \citep{Gaskell+Harrington18}.  

\subsection{Comparisons of lags for different lines from the same ion at the same epoch in the same object}

The complications of off-axis variability and partial obscuration can be avoided by considering lags from similar lines in the same object.  Because the BLR shows substantial radial stratification by ionization (see \citealt{Gaskell09}), it is important to compare the lags of lines with similar levels arising from the same ion. The most practical lines for doing this are the Balmer lines because they are strong, relatively unblended lines that arise from very nearly the same gas and have convenient wavelengths for ground-based observations.  It has long been known that H$\beta$ is broader than H$\alpha$.  \citet{Osterbrock77} found the FWHM of H$\beta$ to be $20 \pm 4$\% broader than H$\alpha$ on average in his sample of 36 Seyfert 1 galaxies.  Since the lag of a line is proportional to the square of the line width \citep{Krolik+91} this means that the effective radius of H$\alpha$ emission is $44 \pm 8$\% greater than the effective radius of H$\beta$ emission.  The physical interpretation of this is that the Balmer decrement is flatter at smaller radii \citep{Shuder82}.  This is also predicted by the self-shielding photoionization model of the BLR \citep{Gaskell+07}.  \citet{Zu+11} and \citet{Peterson+04} give lags for at least two of H$\alpha$, H$\beta$, and H$\gamma$ for 11 PG quasars.  Using the lags from the GS method the lags of H$\alpha$ are, on average, $32 \pm 11$\% larger than those of H$\beta$, in agreement with expectations.  For H$\beta$ and H$\gamma$ there is no significant difference in the lags.

After reducing the H$\alpha$ lags by a factor of 1.3, the $\chi^2$/~$dof$ of the differences in lags between H$\alpha$, H$\beta$, and H$\gamma$ is 0.43 using the published smoothed-bootstrap errors.  However, if the smooth bootstrap errors are reduced by a factor of 1.36 as advocated above, $\chi^2$/~$dof \approx 0.83$, which is not significantly different from unity and thus supports the appropriateness of the smooth bootstrap errors after re-scaling.

A similar analysis using the JAVELIN lags given by \citet{Zu+11} for the same objects gives $\chi^2$/~$dof = 13$. To obtain $\chi^2$/~$dof \approx 1$ the JAVELIN error bars need to be increased by a factor of 3.6 with a 68\% confidence interval of (3.3, 4.7).  A similar scale factor is obtained if one considers just the single-line fits (groups A and B of \citealt{Zu+11}) or just the multiple line fits (groups C and D - see \citealt{Zu+11} for an explanation of their process of simultaneously fitting more than one line).

\section{What is JAVELIN measuring?}

Since NGC~5548 has been very well studied by reverberation mapping, it is particularly useful for intercomparing lag estimates from different methods.  Since the two methods with smallest errors according to the published error estimates are the response function fitting method of Cackett \& Horne and JAVELIN, we would expect JAVELIN to show the best agreement with the Cackett \& Horne estimates.  In fact, this is {\em not} the case.  Figure 3 shows that, surprisingly, the JAVELIN lags correlate best with the {\em peak} lags determined by the GS method.  The difference in the dispersion has a two-tailed significance of $p \approx 0.01$.  This suggests that what JAVELIN is measuring is most closely related to the {\em peak} of the CCF, rather than the centroid.  Comparison of the JAVELIN lag estimates for the other objects in Table 1 of \citet{Zu+11} (objects that were generally not as well observed as NGC~5548) also shows a better correlation with the peak estimates than the centroid, but this is not statistically significant.   If JAVELIN lags are best correlated with the peak lags in the CCF this explains why the errors of the JAVELIN estimates seem to be larger than the error estimates for the centroid using the GS method since the peak has larger errors than the centroid (see Section 3).

 % Figure 3
 \begin{figure}
   \centering \includegraphics[width=11cm]{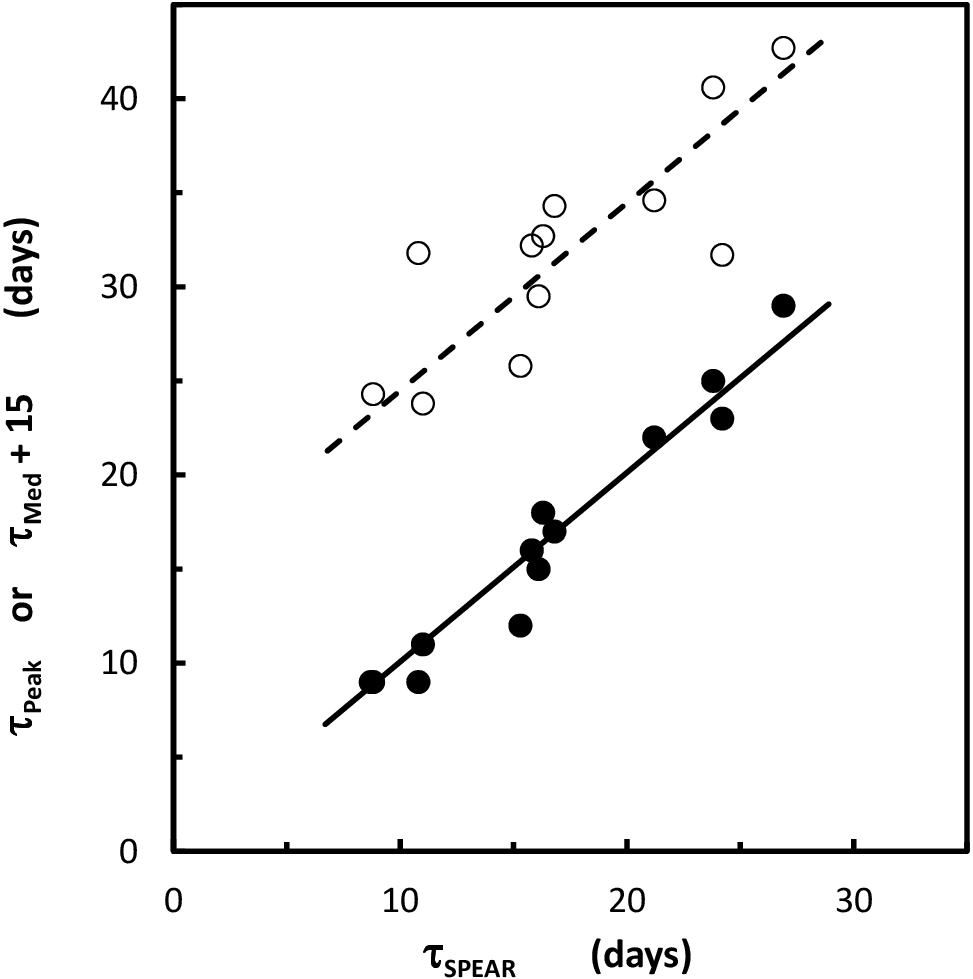}
      \caption{Correlations between the estimates of H$\beta$ lags given by SPEAR/JAVELIN \citep{Zu+11} and estimates from other methods for the well-studied AGN NGC~5548.  The open circles correspond to the lags, $\tau_{med}$, given by the median of the response function \citep{Cackett+Horne06}.  The filled circles correspond to the lags, $\tau_{Peak}$, given by the peak of the CCF using the \citet{Gaskell+Sparke86} method \citep{Peterson+04}.  The $\tau_{Median}$ values have been offset by 15 days for plotting convenience.   The diagonal lines correspond to the lags being equal.  It can be seen that the lags given by JAVELIN correlate best with the {\em peaks} of the cross-correlation function.}.	
       \label{Fig3}
  \end{figure}

\section{The method of \citet{Li+13}}

\citet{Li+13} have also introduced a method which uses a damped random walk to interpolate between observations and also estimates errors, $e_{Li}$, using a MCMC approach, but their method makes different assumptions about the response function.  \citet{Li+13} claim that their method gives effective radii of the H$\beta$-emitting region that are 20\% larger than given by the GS method, but this is a consequence of using the arithmetic mean of ratios to make the comparison.  The arithmetic mean of $\tau_{Li} / \tau_{cent}$, the ratios of lags estimated by \citet{Li+13} to those estimated from the centroid of the CCF using the GS method, is indeed $\approx 1.2$ but the {\em geometric} mean shows that $<\tau_{Li} / \tau_{GS}>$ is statistically indistinguishable from unity.\footnote{The arithmetic mean of the ratio of any set of random positive numbers with the same population means and similar dispersions is greater than unity. Thus, if one looks at the arithmetic mean of the {\em inverse} ratios, $\tau_{cent} / \tau_{Li}$, this would imply that it is the GS method that is giving significantly larger effective radii which is the opposite result to what \citet{Li+13} give.}

For 52 error estimates in common with \citet{Peterson+04} the geometric mean of the ratio of $e_{cent}/e_{Li}$ is $1.8 \pm 0.1$.  Thus, \citet{Li+13} are also claiming a substantial improvement over the GS method, although not as large an improvement as the factor of $\sim 4$ implied by the errors given by \citet{Zu+11} (see section 5 above).  Performing a similar comparison between $e_{cent}$ and $e_{Li}$ to that performed in Section 5 for $e_{JAVELIN}$ gives a $\chi^2 / dof = 3.25$.  To obtain $\chi^2 / dof = 1$ the error estimates of \citet{Li+13} need to be increased by a factor of $4.6 \pm 1.0$.  This suggests that, like JAVELIN, the related method of \citet{Li+13} is also underestimating errors by a substantial factor (albeit less than for JAVELIN).  This means that the errors in the BLR model parameters deduced by \citet{Li+13} are probably also substantially underestimated but analysis of this is beyond the scope of this paper.

\section{Problems with Markov chain Monte Carlo methods}

Y. Zu (private communication) comments that the error report given by JAVELIN is only statistical and that error bars must be underestimated because of the unknown statistics associated with the model assumptions.  Both JAVELIN and the method of \citet{Li+13} use a MCMC method to estimate the errors bars.  A major problem with this is the very large amount of computer time needed to obtain the result.  As has been discussed above, the GS method has been subject to extensive testing using Monte Carlo simulations and the smooth bootstrap method.  To obtain a single error bar by either of these methods typically involves running $\sim 1000$ simulations.  Doing such a large number of runs of JAVELIN for each set of observations is prohibitively computationally expensive at present.  Examining the reasons why the errors in JAVELIN are underestimated is beyond the scope of this paper, but it is perhaps worth mentioning well known problems with MCMC methods.  As \citet{Kelly+14} explain, ``The likelihood function often contains multiple modes ... This presents a difficulty for many optimizers and MCMC samplers.'' and ``there is no guarantee that [the optimization algorithm] will find the global optimum.''  The analysis presented here argues that independent checks of error estimates from MCMC methods are worthwhile and that more development and testing is needed of methods such as JAVELIN to assure that the error estimates given reflect the actual accuracy.

\section{An analytic formula for the error in the lag}

Whilst the analytic formula of \citet{Gaskell+Peterson87} for estimating the error in the lag determined by the GS method (equation 4 in \citealt{Gaskell+Peterson87}) was shown by \citet{Maoz+Netzer89}, \citet{Koratkar+Gaskell91}, and \citet{Oknyanskij93} to agree with Monte Carlo simulations for well-sampled observations, it was also found to grossly underestimate errors when the light curves are poorly sampled and the data are noisy.  As a result of this, the Gaskell \& Peterson analytic formula fell out of regular use after the early 1990s.  However, because of better planning of observations (since lags can be guessed in advance from the radius-luminosity relationship), detector improvements, and more observational resources being devoted to reverberation mapping, essentially all reverberation mapping campaigns {\em are} now well sampled with an observational cadence less than the expected lag.  It is therefore worthwhile to investigate the accuracy of the analytic formula of \citet{Gaskell+Peterson87}.

\begin{figure}
   \centering \includegraphics[width=11cm]{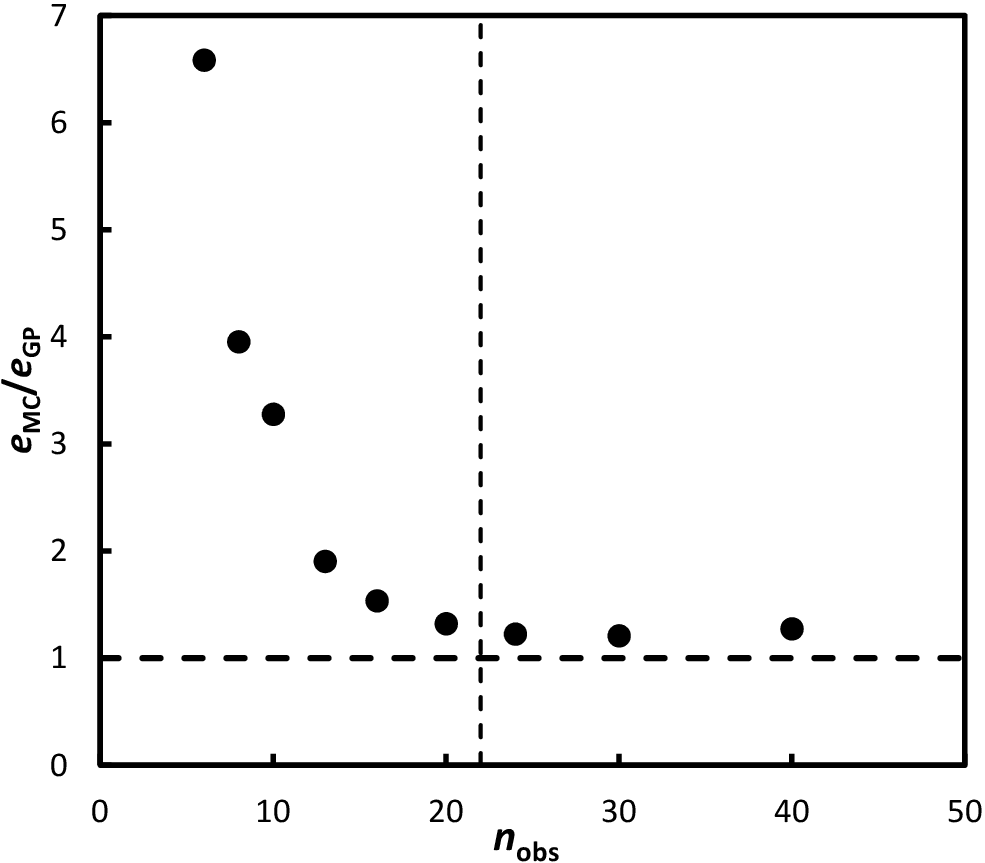}
       \caption{The ratio of the error estimated from the Monte Carlo simulations, $e_{MC}$, to the error calculated from Equation 4 of \citet{Gaskell+Peterson87}, $e_{GP}$, as a function of the number of observations, $n_{obs}$. The points are the averages of the simulations using artificial light curves in Tables 1 and 2 of \citet{White+Peterson94}. The horizontal dashed line is $e_{MC} = e_{GP}$ and the vertical line shows
       approximately where the simulations indicate that the formula of \citet{Gaskell+Peterson87} starts to significantly underestimate the error.  It can be seen that, except for ill-sampled light curves, the errors given by the Monte Carlo simulations are close to the errors given by the analytic formula of \citet{Gaskell+Peterson87}.}.		
       \label{Fig4}
 \end{figure}

\citet{White+Peterson94} give both the errors from Eq.~4 of \citet{Gaskell+Peterson87} and from Monte Carlo simulations for artificial light curves as the number of observations is reduced.  The ratio is plotted in Fig.~\ref{Fig4} as a function of the number of observations.  The ratio of the cadence of the observations to the lag varies from 0.25 in the best case (right-hand side of the figure) to 2.0 in the worst case (left-hand side of the figure).  The simulations suggest that, for well-sampled light curves, the \citet{Gaskell+Peterson87} formula is only underestimating the errors by a factor of $\sim 1.3$ compared with the Monte Carlo simulations.

% Figure 4
 % This figure comes from RSSFR_testing_02 Koratkar+Gaskell

\begin{figure}
   \centering \includegraphics[width=11
   cm]{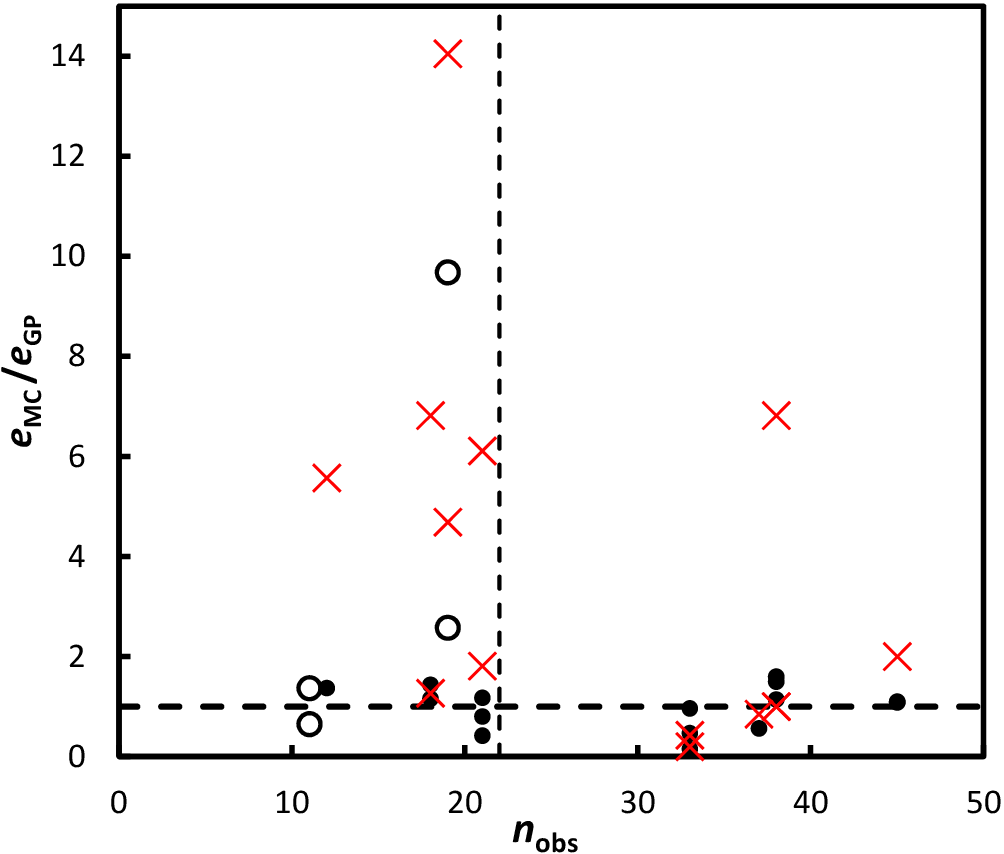}
      \caption{The ratio of the error estimated from the Monte Carlo simulations, $e_{MC}$, to the error calculated from Equation 4 of \citet{Gaskell+Peterson87}, $e_{GP}$, as a function of the number of observations from the analysis of  \citet{Koratkar+Gaskell91}.  Solid
      circles are for observations of strong lines (Ly$\alpha$, C~IV, C~III], or Mg~II) with a median cadence of less than twice the lag.  The open
      circles are for strong lines with a median cadence of more than twice the lag.  The red crosses are weak lines or parts of lines.  As in Figure 4, the horizontal dashed line is $e_{MC} = e_{GP}$ and the vertical line shows approximately where the simulations of \citet{White+Peterson94} indicate that the formula of \citet{Gaskell+Peterson87} starts to significantly underestimate the error.  It can be seen that for the strong lines with the best sampling the errors given by the Monte Carlo simulations are, on average, similar to the errors given by the Gaskell \& Peterson formula.}.	
       \label{Fig5}
 \end{figure}

For their sample of reverberation-mapped AGNs \citet{Koratkar+Gaskell91} give both errors calculated from Eq.~4 of \citet{Gaskell+Peterson87} and estimated from Monte Carlo simulations.  For convenience, these are plotted in Fig.~\ref{Fig5} as a function of the number of observations, but it should be recognized that the ratio of the cadence of the observations to the lag is larger (i.e., worse) than for the \citet{White+Peterson94} simulations.  Also, for the weaker features the flux errors are larger because the spectra were obtained with the {\it IUE} satellite.  These factors have been indicated through the use of different symbols (see figure caption).  If one looks only at the best-sampled observations for the strongest features (the solid circles in Fig.~\ref{Fig5}), it can be seen that the \citet{Gaskell+Peterson87} formula is not systematically underestimating the errors.

 % Figure 5
% This figure comes from RSSFR_testing_02 Koratkar+Gaskell

\begin{figure}
   \centering \includegraphics[width=8.5cm]{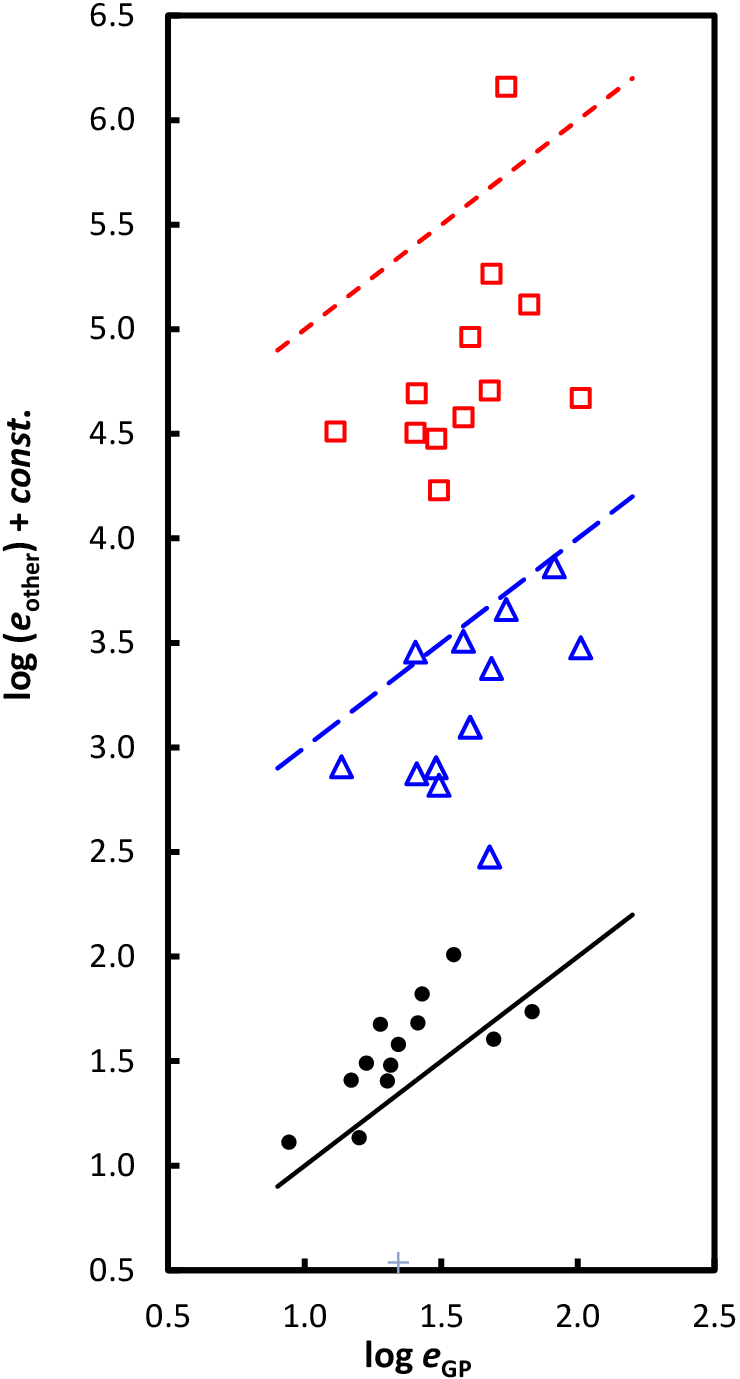}
      \caption{Errors, $e_{GP}$, given by the Gaskell \& Peterson formula compared with errors determined by other means for the H$\beta$ and continuum observations of \citet{Kaspi+00}.  
      %The abscissa in each case is $e_{GP}$, the error given by Eq.~4 of \citet{Gaskell+Peterson87} for the lags estimated from the centroid of the CCF calculated by the \citet{Gaskell+Sparke86} method.  
      Solid black circles show the relationship for $e_{bootstrap}$, the uncorrected smoothed bootstrap errors of \citet{Peterson+04} for the GS method.  The blue triangles show the relationship between $e_{GP}$ and $e_{Li}$, from  \citet{Li+13}.  Red squares are for $e_{GP}$ and $e_{JAVELIN}$. Blue triangles and red squares are offset by 2 and 4 dex respectively for plotting convenience.  Diagonal lines indicate perfect agreement.}.	
       \label{Fig6}
 \end{figure}

It is important to recognize that the pioneering {\it International Ultraviolet Explorer} satellite ({\it IUE}) observations analyzed by \citet{Koratkar+Gaskell91} are much worse than modern reverberation mapping.  The {\it IUE} spectra are noisy, most of the observations were not made with reverberation mapping in mind, and if they were, reverberation times were thought to be much longer than we now realize they are.  Given that the \citet{Gaskell+Peterson87} formula is agreeing with the Monte Carlo error estimates for the solid points in Fig.~\ref{Fig5}, we can therefore expect that it will do much better for modern observing campaigns.  The observations of \citet{Kaspi+00} are more typical of modern reverberation mapping campaigns, although campaigns doing detailed velocity-resolved reverberation mapping (e.g., \citealt{Sergeev+99}; \citealt{Kollatschny+01}) have even better sampling.  For the \citet{Kaspi+00} observations, Fig.~\ref{Fig6} shows comparisons of $e_{GP}$, the lag errors given by the \citet{Gaskell+Peterson87} formula, versus other error estimates.  The three other methods shown are the smooth bootstrapping method of \citet{Peterson+98} (after correction by a factor of 1.36 discussed in Section 2), JAVELIN, and the method of \citet{Li+13}.

% Figure 6
 % This figure comes from RSSFR_testing_02 GP-Kaspi

There are several things to notice from Fig.~\ref{Fig6}.
\begin{enumerate}
\item There is good agreement between the \citet{Gaskell+Peterson87} formula and the bootstrap error estimates.
\item There is poor agreement with the error estimates of JAVELIN and \citet{Li+13}.  
\item The JAVELIN method and the method of \citet{Li+13} are systematically giving error estimates that are too small by a large factor, as has been discussed above.
\item There is a much smaller offset of $0.19 \pm 0.05$ dex between the error given by the \citet{Gaskell+Peterson87} formula and the error given by the smooth bootstrap method.
\item The correlations between $e_{JAVELIN}$ and $e_{GP}$ and between $e_{Li}$ and $e_{GP}$ are poorer than the correlation between the bootstrap error estimates and $e_{GP}$
\end{enumerate}

The last of these suggests that the \citet{Gaskell+Peterson87} formula is underestimating the true errors by a factor of $1.55 \pm 0.12$. This factor is similar to the factor of 1.3 suggested above by the simulations of \cite{White+Peterson94} when their artificial light curves are well sampled.  Note that, if allowance is made for this probable factor of 1.3 to 1.5 underestimate of $e_{GP}$, then the systematic differences between the errors given by JAVELIN and \citet{Li+13} are even larger than implied by Fig.~\ref{Fig6}.

The scatter between the bootstrap error estimates and $e_{GP}$ is 0.18 dex while the scatters between $e_{GP}$ and the error estimates from JAVELIN and \citet{Li+13} are 0.46 and 0.35 dex respectively.  The increase in scatter in the comparisons with the latter two methods are significant ($p = 0.002$ and 0.02 respectively).  The scatter between the estimates of JAVELIN and the method of \citet{Li+13} for the data of these observing campaigns is 0.49 dex, which is again significantly greater than the scatter between the bootstrap error estimates and $e_{GP}$ ($p = 0.001$).  In the simulations of \citet{Peterson+98} the scatter between the bootstrap error estimates and Monte Carlo estimates is 0.23 dex for the upper and lower limits treated separately or 0.13 dex if they are averaged.  Since the dispersions in the sets of errors \citet{Peterson+98} obtain from their smooth bootstrap method and their Monte Carlo simulations are statistically indistinguishable ($p = 0.5$) after the former have been reduced by 1.36, the errors in each method contribute equally to the scatter between the two methods.  We can thus conclude from the the scatter between the bootstrap error estimates and $e_{GP}$ (after increasing the latter by a factor of 1.45) that, for well-sampled data sets, {\em the \citet{Gaskell+Peterson87} formula is giving error estimates which are as reliable as those given by Monte Carlo simulations or the smooth bootstrap method.}

With the heuristic correction implied by the comparisons given above with both Monte Carlo simulations and the smooth bootstrap method of \citet{Peterson+98} using both artificial and real data we get the following equation for the error:

\begin{eqnarray}
\qquad\qquad\qquad\quad e = \frac{0.55 \, W_{CCF}}{1 + r_{peak}(n - 2)^{1/2}}
\end{eqnarray}

\noindent where $r_{peak}$ is the height of the peak in the cross-correlation function, $W_{CCF}$ is the full width at half maximum (FWHM) of the peak, and $n$ is the number of data points.\footnote{Note that the notation of Eq.~1 here differs slightly from Eq.~4 of \citet{Gaskell+Peterson87} in that $W$ now refers to the {\em full} width at half maximum.}

\section{Planning observing campaigns}

Eq.~1 can be used to estimate in advance what accuracy will be achieved by an observing campaign and thus to optimize the campaign to achieve the desired results while being good stewards of the observational resources.  It can be seen from Eq.~1 that $e$ is insensitive to $r$.  Furthermore, inspection of correlation functions shows that $r$ is generally about 0.8 with relatively little dispersion for good data.   $e$ thus depends mainly on $W$ and $n$.  So long as the sampling is adequate (i.e., it fulfills the Nyquist criterion) $W$ depends on the apparent characteristic timescale of variability of the AGN.  To get a useful CCF one needs to observe at least from one trough in the continuum light curve to the trough in the line light curve corresponding to the next continuum trough (or a similar duration relative to peaks in the two time series rather than troughs).  This means observing for $\gtrsim 2W$.  Since one does not know in advance when peaks or troughs will occur, it is safer to observe for $\sim 4W$.  The minimum duration of the observing campaign (the observing window) thus depends on $W$.  $W$ is also the dominant factor in determining $e$ in Eq.~1.  It is thus very helpful for planning both the duration of an observing campaign and the cadence of observations to have an idea of what $W$ is in advance.

\begin{figure}
   \centering \includegraphics[width=11cm]{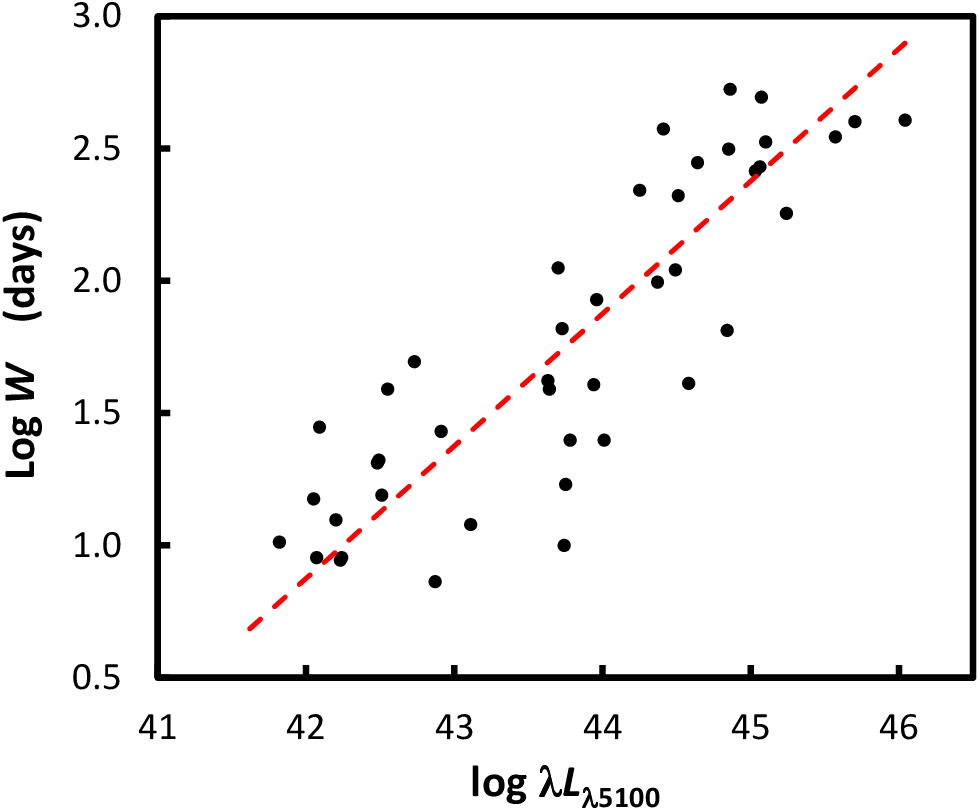}
      \caption{The full width at half maximum, $W$, of the H$\beta$ and optical continuum cross-correlation function versus the luminosity, $\lambda L_{\lambda 5100}$, for reverberation-mapped AGNs (see text).  The dashed red line shows the expected theoretical relationship $W \propto L^{1/2}$.}.	\label{Fig7}
 \end{figure}

Physically, a characteristic timescale comes from a size of an emitting region and the appropriate velocity for the propagation of changes.   Given that the spectral energy distribution of AGNs in the optical and UV is very similar after allowance for host galaxy light and reddening \citep{Gaskell+04,Heard+Gaskell23} the radial temperature structure of the accretion disc in AGNs is very similar from object to object (see \citealt{Gaskell08}).  The radius contributing the most to a given monochromatic flux thus depends only on $L^{1/2}$.  If we assume that the velocity of propagation of the ill-understood mechanism responsible for changes in the continuum emission is the same in all AGNs, then we expect $W \propto L^{1/2}$ as well.  Fig.~\ref{Fig7} shows the correlation between $W$ and the optical luminosity, $\lambda L_{\lambda 5100}$, measured at $\lambda$5100 for AGNs with good reverberation mapping.  Luminosities have been taken from Table 13 of \citet{Bentz+13} and $W$ has been measured from the relevant observational papers (see footnotes to Table 12 of \citealt{Bentz+13} for references to the observational papers).  As can be seen, there is a good correlation and the slope is in good agreement with the expected $W \propto L^{1/2}$.  The scatter (0.3 dex) is similar to the scatter in the radius--luminosity relationship (see Table 14 of \citealt{Bentz+13}).  Since the lag of H$\beta$, $\tau_{H\beta}$, is also proportional to $L^{1/2}$ this means that $W \propto \tau_{H\beta}$.  The scatter in this relationship is $\pm 0.23$ dex and $W \thickapprox 2.5 \tau_{H\beta}$.

% Figure 7
 % This figure comes from ACF-luminosity_02 lag-error_paper

From Fig.~7 we get
\begin{eqnarray}
\qquad\qquad\qquad \log W = 0.5 \log \lambda L_{\lambda 5100} - 20.18
\end{eqnarray}

While Eq.~2 can be used to estimate $\log W$ to $\pm 0.3$ for the purposes of planning observing campaigns, the reader is cautioned against trying to make physical deductions from Eq.~2.  This is because the optical variability of an AGN can be approximated as a damped random walk \citep{Gaskell+Peterson87} and, as was pointed out by \citet{Welsh99}, with a red noise power spectrum {\it the width of the ACF is strongly influenced by the length of the observing window.}  Reverberation mapping observations designed for studying broad lines such as the Balmer lines are particularly poorly suited for determining continuum variability timescales because the length of the observing window is often chosen to match the expected emission line reverberation time, which is also proportional to $L^{1/2}$.  A relationship similar to that in Fig.~7 can readily be reproduced simply by sampling a damped random walk time series with different length windows.  However, the degree to which there is a physical basis to the relationship in Fig.~7 does not matter for the purposes of using the predicted $W$ for planning observations.  What matters in calculating $e$ is the {\em observed} $W$ and observing campaigns will probably be reproducing the conditions similar to the observations giving Fig.~7.

What might seem at first to be a surprising consequence of $W$ being linearly proportional to the length of the observing window for red noise is that, according to Eq.~1, a shorter observing window can give a more accurate lag for the same number of observations.  A reduction in the length of the observing window can obviously be made after the observations have been taken simply by considering subsets of the observations and averaging the results.  This is effectively what is being done when the data are de-trended before cross-correlating.  As \citet{Welsh99} has discussed, de-trending produces a gain in accuracy in the determination of the lag.

At this point, two {\em major} cautions are needed:

\begin{enumerate}
\item As discussed above (see Section 6), the lag in an object changes on the timescale of continuum events \citep{Gaskell+21}. To get reliable size estimates, an observing campaign needs to catch {\em multiple} continuum events.  Once one has good enough sampling for the lag one is trying to measure, the best use of observational resources is to observe {\em longer}, not more frequently.
\item Observing campaigns that are short compare with the lag can significantly underestimate the lag \citep{Welsh99}.   For example, \citet{Gaskell+22} have shown that this was the source of conflicting reverberation mapping sizes for the region emitting Fe\,II. The poorer quality campaigns were giving systematically too small Fe\,II lags, by around a factor of two.
\end{enumerate}

Unfortunately, non-astronomical constraints often lead to shorter-than-desired monitoring campaigns.

\section{The discrete correlation function}

For completeness, brief mention should be made of a much less powerful approach used in AGN research to determining lags from irregularly sampled time series: the discrete correlation function (DCF).  This method, which was first proposed by \citet{Mayo+74} and \citet{Mayo78}, is best known in astronomy through the advocacy of \citet{Edelson+Krolik88}.  Important corrections to the methodology of \citet{Edelson+Krolik88} are given in Section 2.2 of \citet{White+Peterson94}.\footnote{The apparently large differences between the \citet{Gaskell+Sparke86} method and the \citet{Edelson+Krolik88} method in calculating the H$\beta$ versus optical continuum cross-correlation function for NGC~5548 in \citet{Peterson+92} (see their Fig.~4) are a result of problems in the method as originally described by \citet{Edelson+Krolik88}.  For a better comparison using the same data, see Fig.~1 of \citet{White+Peterson94}.}  Further refinements have been introduced by \citet{Alexander97,Alexander13}.  The DCF is an example of what is known in the statistical literature as a slotting technique (see \citealt{Babu+Stoica10} and \citealt{Rehfeld+11} for details).  Rather than using interpolation in the data domain, the correlation function is determined by binning products of corresponding data points from the two time series in the lag domain, so that observations only contribute to the correlation function in a given lag bin if their lagged time difference deviates by less than half the lag bin width.  More generally, observations can be weighted by a kernel.  The DCF method uses a rectangular kernel (i.e., the weighting of a product is 1 if it falls within half the bin width and 0 if it fall outside this interval.)  Other kernels can be used and offer some improvements (see discussion in \citealt{Rehfeld+11}).  The advantages claimed by \citet{Edelson+Krolik88} for the DCF method over the GS method are that it provides an assumption-free representation of the correlation function and that it ``allows meaningful error estimates.''  A problem sometimes encountered in reverberation mapping is the issue of correlated errors (see \citealt{Gaskell+Peterson87}) where random errors in one time series are correlated or anti-correlated with the random errors in the other time series at a given epoch.  If necessary, points at zero lag can easily be removed when calculating the DCF \citep{Edelson+Krolik88}, while in the GS method the effects of correlated errors on the lag need to dealt with through modelling (see section IV.d of \citealt{Gaskell+Peterson87}.)  However, while it is always worthwhile remembering the possibility of correlated errors, in practice, in modern reverberation mapping campaigns the effect of correlated errors is very small and, when the sampling is good, they are very obvious.  The CCFs calculated by the GS method for five AGNs shown Fig.\@14 of \citet{Peterson+98} provides a good illustration of this.  Two of the five show small negative blips at zero correlation implying anti-correlated errors and the other three show small positive blips implying correlated errors.  As can be seen these blips are very small and narrow and they will not have any effect on the lag estimate.

\begin{figure}
   \centering \includegraphics[width=11cm]{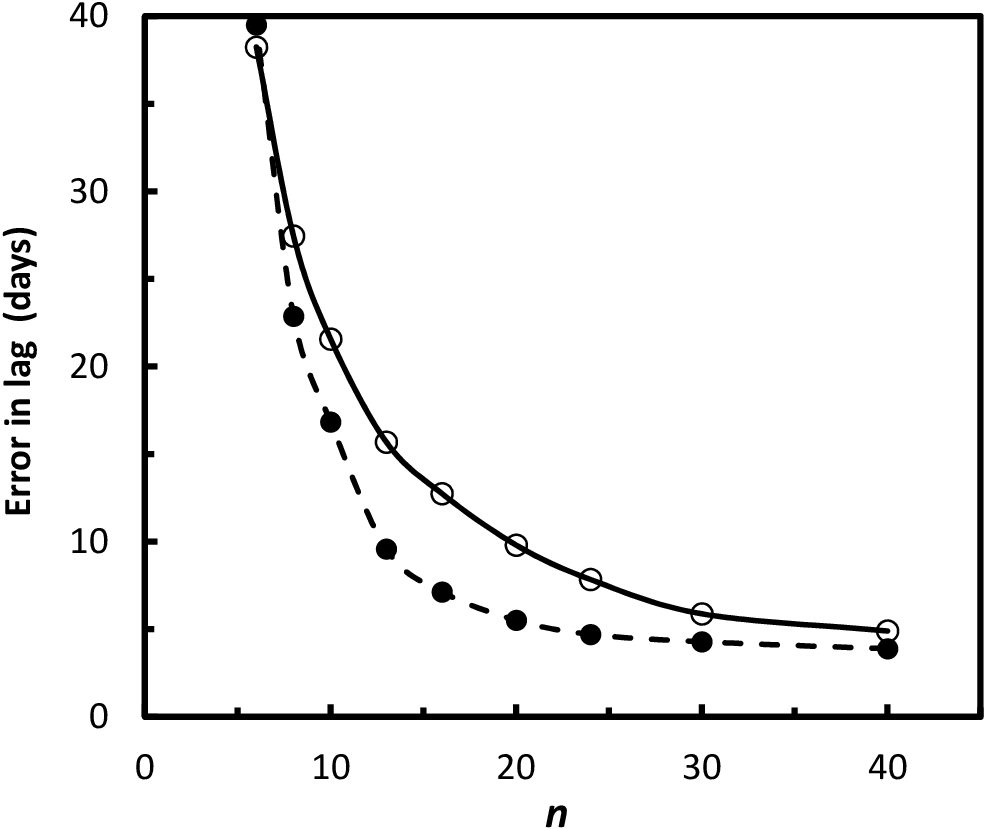}
      \caption{Monte Carlo estimates of the errors in the lag for an artificial continuum convolved with the response function of a BLR of radius 20 light
      days as a function of the number of observations.  40 observations corresponds to a mean sampling interval of 5 days (i.e., $\sim$ a quarter of the light travel time).  The filled circles are for the lag calculated using the \citet{Gaskell+Sparke86} method while the open circles use the discrete correlation method following the prescription of \citet{White+Peterson94}.   It can be seen that while both methods are almost equally good for the best sampling sampling and are equally bad for very sparse sampling, the DCF method gives error bars that can be $\sim 2 \times$ larger for intermediate cases.   Data are taken from Tables 1 and 2 of \citet{White+Peterson94}.}.	\label{Fig8}
 \end{figure}

For well-sampled data sets the difference between the cross-correlation function determined by the GS method and the DCF method is not a problem (see, for example, Fig.~1 of \citealt{White+Peterson94}), but, if the time series are less well sampled, the DCF looks {\em much} worse (see, for example, Figs.~5 -- 7 of \citealt{Edelson+Krolik88}).  \citet{Edri+12} note that the ZDCF of \citet{Alexander97} also produces more erratic results than the GS method (see their Fig.~6). It is easy to show that these problems arise is because the DCF method a statistically less powerful method (i.e., more prone to give Type II errors).  For example, as a demonstration of this one can simply remove points from a well-sampled time series and see that the GS method continues to give a clear peak in the cross-correlation function and hence an estimate of the lag, while the discrete correlation function does not.  An illustration of this is given in Fig.~3 of \citet{Peterson93}.  Monte Carlo simulations of \citet{White+Peterson94} using artificial light curves show that the GS method gives {\em substantially} smaller error bars than the DCF method except for very good (or very bad) sampling.  As can be seen in Fig.~\ref{Fig8}, there is {\em a difference of a factor of two} in the lag errors for the intermediate case.  The study of \citet{Litchfield+95} gives the same result. For the case they consider with typical sampling (their Table 2) they get errors for the DCF method which are a factor of $2.1 \pm 0.2$ worse than the GS method.  The probability distributions can be compared in their Fig.~5.  \citet{Litchfield+95} also consider ``eye estimates'' (cf. \citealt{Clavel+87}) where the lag is simply estimated by comparing the positions of peaks in the two light curves. They find that {\em the DCF method give errors in the lag which are a factor of $1.6 \pm 0.1$ worse than the eye estimates.}

% Figure 8
 % This figure comes from RSSFR_testing DCF

The comparison in Fig.~\ref{Fig8} means that the DCF method is statistically {\em much} less powerful than the GS method because more observations (often around four times as many) are needed to obtain a comparable accuracy.

One of the alleged advantages of the DCF method is that it is claimed to be straightforward to define an error estimate for each bin (see Eq.~5 of \citealt{Edelson+Krolik88}). It has been the universal practice to plot the DCF as a series of discrete points each with its own error bar.  Often these points are plotted over the cross correlation function derived with the GS method.  If the number of normalized products in a bin is small, the error bar will be large.  These error bars are assumed to be giving the independent errors applicable to the range of the correlation function covered by each bin.  However, inspection of a typical DCF from well-sampled monitoring such as the DCF shown in Fig.~9 shows that {\em the point-to-point scatter is much less than the error bars given by the DCF method}.  Fig.~10 shows the residuals from the polynomial fits to the two sides of the DCF.  It can be seen that a Gaussian with a width corresponding to the error bars given by the DCF method is a poor fit to the residuals. In general, the scatter is about a factor of two less than the error bars given by the DCF method.  The smaller Gaussian corresponds to the spread covered by 67\% of the scatter, but it can be seen that the distribution is clearly non-Gaussian with some of the outlying points deviating by an amount consistent with the DCF error bars.  The sort of behaviour shown in Figs.~8 and 9 is common (for some additional examples see Fig.~7 of \citealt{Reichert+94}, Fig.~5 of \citealt{Rodriguez-Pascual+97}, and Fig.~7 of \citealt{O'Brien+98}).  This question of the size of error bars for individual lag bins is separate from the problem of the lower accuracy of the DCF method in determining lags since the errors of the individual bins are not used in studying the accuracy of DCF lag estimates.

% Figure 9
\begin{figure}
   \centering \includegraphics[width=11cm]{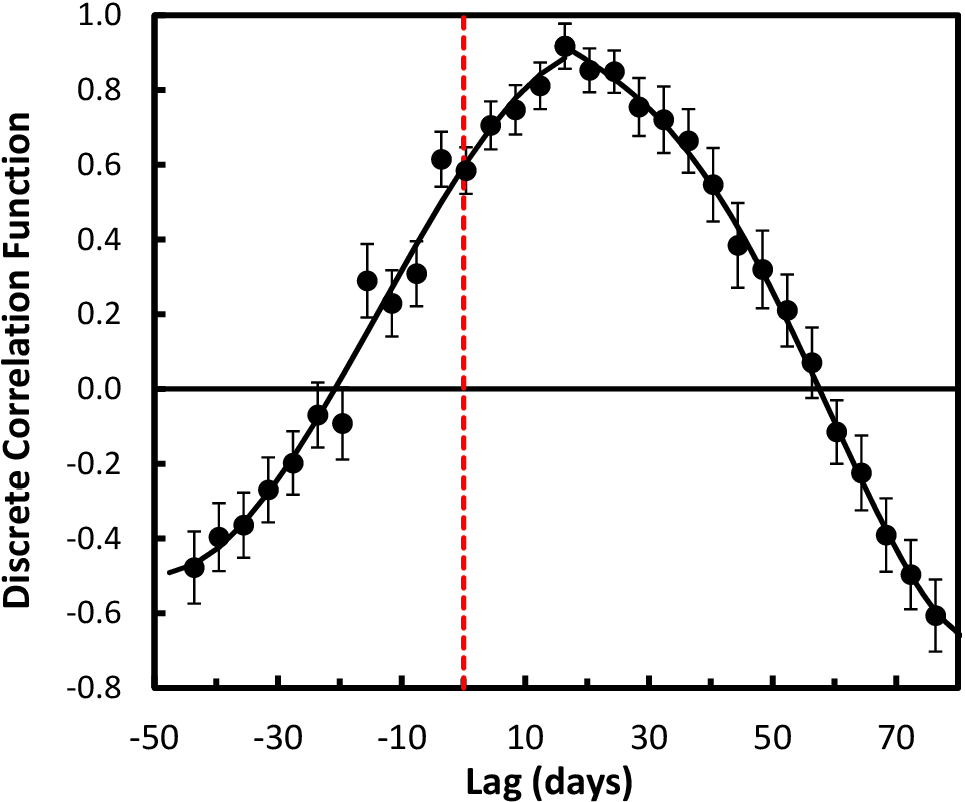}
      \caption{The peak of the discrete correlation function for H$\beta$ lagging the optical continuum at $\lambda$5100 for NGC~5548
      as observed in 1989--1990 by \citet{Peterson+92}.  The DCF is calculated with the corrections of \citet{White+Peterson94}.  The two solid
      curves are polynomial fits to the DCF on either side of the peak.  It can be seen that the points agree with the smooth curves far better than
      would be expected with the error bars given by the DCF method.}.	
       \label{Fig9}
 \end{figure}
 % This figure comes from CCF_DCF_comparisons

% Figure 10
\begin{figure}
   \centering \includegraphics[width=10cm]{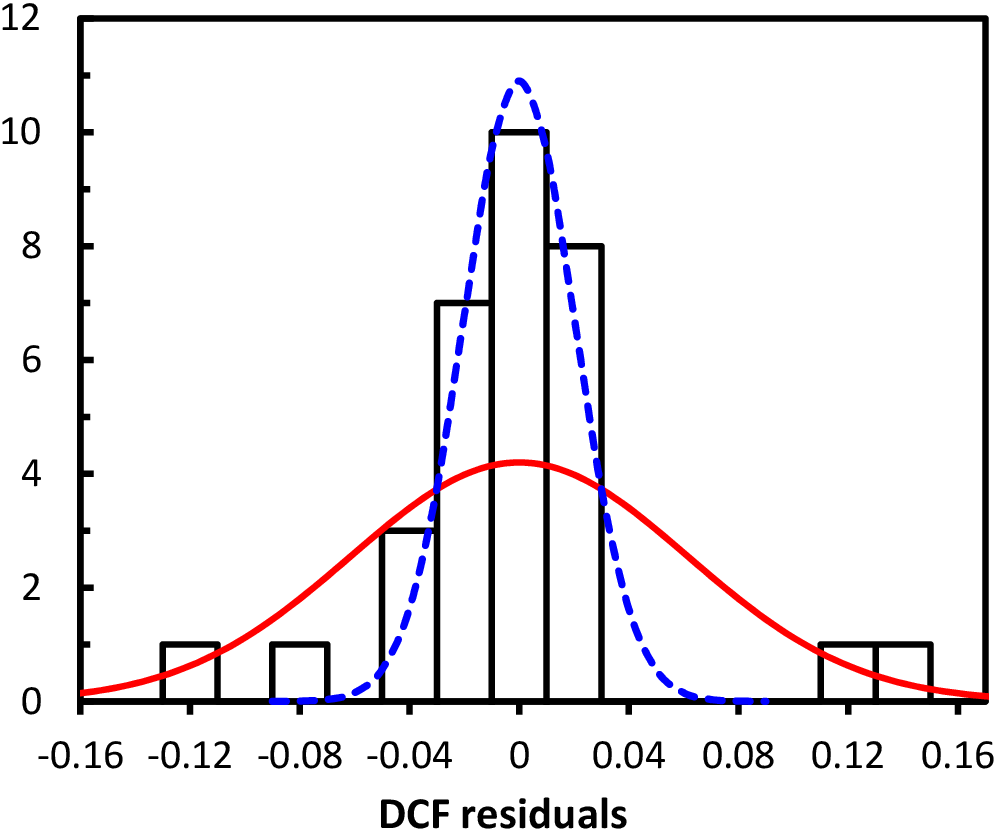}
      \caption{The distribution of residuals of the discrete correlations in Fig.~9 from the polynomial fits.  The solid red curve is a Gaussian with a width
      corresponding to the mean of the errors given by the discrete correlation function method (i.e., the error bars in Fig.~9) fit to the distribution.  The dashed
      blue curve is a Gaussian with a width set from the spread covered by 67\% of the scatter.}.	
       \label{Fig10}
 \end{figure}
 % This figure comes from CCF_DCF_comparisons

The reason why the DCF is not as powerful as the GS method is that the former weights the result according to when the observations were taken, whilst the latter weights the whole light curve more evenly \citep{Gaskell+89}.  The problem with weighting according to when the observations were taken is that these times might not correspond to when the AGN was varying.  As \citet{Gaskell+89} explained, these problems get hidden by the binning of the DCF.  They are revealed by looking at the {\em un}binned DCF. \citet{Koratkar+Gaskell89} show a clear example of this in their Fig.~14 (reproduced here as Fig.~11)\footnote{Note that an un-binned DCF has to follow the normalization of \citet{Edelson+Krolik88} rather than that of \citet{White+Peterson94}.} It can be seen that there is a lot of structure in the unbinned distribution of cross-products as a function of the phase delay.  This is typically in the form of clouds of points and horizontal bands. The structure in Fig.~11 arises from there being many observations when the flux does not change much.  The simple binning of the DCF method makes no allowance for the clumping of cross-products.

% Figure 11
\begin{figure}
   \centering \includegraphics[width=11cm]{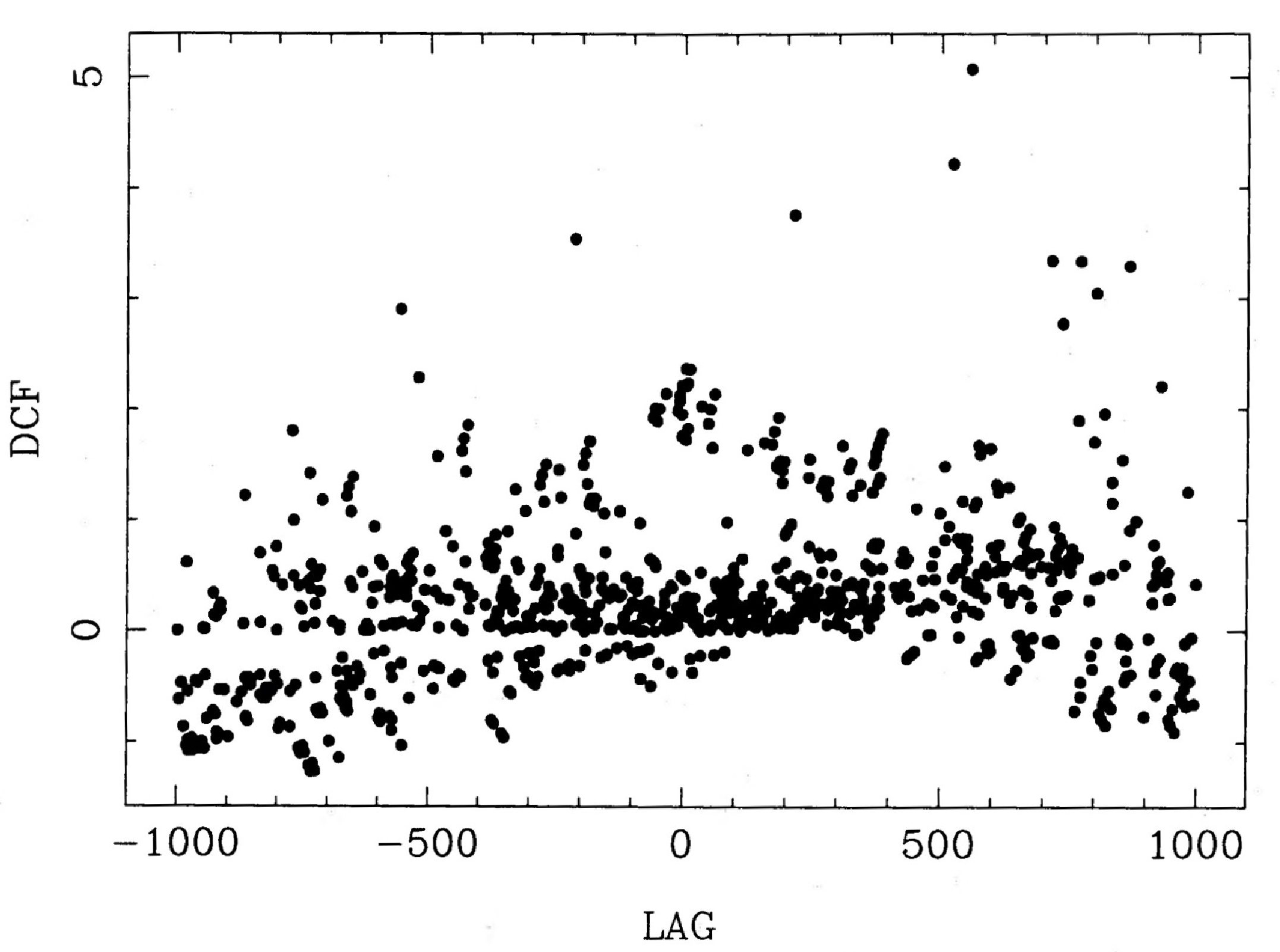}
      \caption{The unbinned DCF of the C~IV line and the UV continuum for Fairall 9 as a function of lag in days.  Periods of more intense sample produce bands and clumps of points.  (Figure reproduced from \citealt{Koratkar+Gaskell89})}.	
       \label{Fig11}
 \end{figure}
 % This figure is from Koratkar & Gaskell (1989)

In this paper we have only been concerned with the accuracy with which lags and their associated errors can be determined for AGNs.  Although the DCF method is often a significantly less powerful method that the GS method for estimating lags, it should be noted that the latter {\em does} have biases in determining the {\em shape} of the auto-correlation function (ACF).  This is explored by \citet{Rehfeld+11}.  An interpolation scheme causes an enhancement of low frequencies relative to high frequencies and this means that the auto-correlation function is somewhat broader in the wings than it should be. This effect is illustrated in Fig.~7c of \citealt{Rehfeld+11}.  In AGN research, however, we are not concerned about the shape of the ACF and for modern, relatively well-sampled data sets the difference in the shape of the ACF is slight. It can be seen to cause a slight systematic difference between the CCFs calculated by the DCF method and by the GS method when the CCF goes to a minimum (see, for example, between -90 and -50 days and between +60 and +110 days for the CCF of NGC~5548 shown in Fig.~1 of \citealt{White+Peterson94}).  As mentioned above, the kernel used by \citet{Edelson+Krolik88} is rectangular. \citet{Rehfeld+11} show that the best results for the shape of the CCF can be obtained using a Gaussian kernel (see their Fig.~7c and 7e).

\section{The Modified Cross-Correlation Function \\(MCCF)}

In the case where one time series is much better sampled than the other (as has often been the case when comparing optical and IR observations - see \citealt{Oknyanskij+99}) it is possible to improve on the DCF method with an approach that has been called the ``modified cross-correlation function'' (MCCF) \citep{Oknyanskij93,Oknyanskij+99} that is intermediate between the GS method and a DCF approach.  The MCCF method imposes a restriction on the use of interpolated data.  It is only used if the separation in time from the nearest observation is less than some fixed width, $w$.  The parameter $w$ is set by taking into consideration the sampling, the variability timescale, and the measurement errors.  For small $w$, the MCCF method produces results similar to the DCF method.  As $m$ increases to half the maximum gap between observed points in the worst-sampled time series the MCCF method becomes the GS method.  A comparison of results from the three methods can be seen in Fig.~2 of \citet{Oknyanskij93}.\footnote{Note that the DCF is calculated using the older normalization of \citet{Edelson+Krolik88} and hence does not match well.}  \citet{Oknyanskij93} and \citet{Oknyanskij+99} find that their MCCF centroid lag errors estimated by Monte Carlo simulations are in agreement with the analytic formula of \citet{Gaskell+Peterson87}.  A {\it Python} version of code for calculation the MCCF is available (see \citealt{Oknyansky+Oknyansky22}).

\section{Conclusions and recommendations}

Analysis of the results of Monte Carlo simulations shows that, for reasons that are easily understood, the smooth bootstrap method of \citet{Peterson+98} systematically overestimates of the errors in the reverberation lag determined by the widely used interpolation method of \citet{Gaskell+Sparke86}. Comparison of lags for lines of the same ion in the same objects supports this conclusion.  It is recommended that error estimates obtained by the bootstrap method be reduced by a factor of 1.36.  Examination of the error estimates given by the analytical formula of \citet{Gaskell+Peterson87} shows that after a modest rescaling (Eq.~1 above) it gives reliable estimates of the error in the lag when the cadence of the observations is $\lesssim$ the lag.  An improved version of this formula is given by Eq.~1 above.  In practice, the width, $W$, of the CCF used in Eq.~1 is proportional to $L^{1/2}$ but the physical basis of this is uncertain because of the strong influence of the length of the observing window.

The overestimate of the errors by the bootstrap method needs to be allowed for when evaluating other methods of lag determination that might offer possible improvements.  In particular, the response-function-fitting method of \citet{Cackett+Horne06} seems only to offer a slight improvement.

Making comparisons of the JAVELIN/SPEAR method of \citet{Zu+11} with the GS method and making comparisons of lags obtained for lines of the same ion in the same objects indicates that JAVELIN {\em substantially} underestimates the errors in its estimated lags.  The reported high accuracy of the JAVELIN method thus seems to be spurious.  Instead, the accuracy of the JAVELIN method seems to be slightly poorer than the standard method of \citet{Gaskell+Sparke86}.  A related method proposed by \citet{Li+13} seems to have similar but less severe problems.  Caution is in order when estimating errors by MCMC methods.

Finally, it has been pointed out that the discrete correlation function method \citep{Edelson+Krolik88} gives errors in the lag which are often {\em twice as great} as those of the \citet{Gaskell+Sparke86} method.  There are also problems with the correlation function error bars given by the DCF method.  The shortcomings of the DCF method for estimating lags have been in the literature for over two decades and deserve to be better known.

In the light of the above, a number of specific recommendations can be made:

\begin{enumerate}

\item The GS interpolation method remains the simplest and one of the best ways to determine the lag.

\item Except for ill-sampled data with a cadence longer than the expected lag, the easiest way to estimate the error in the lag is from Eq.~1.

\item Errors calculated using the smooth bootstrap (FR/RSS) method of \citet{Peterson+98} should be reduced by a factor of 1.36.

\item The discrete correlation function (DCF) method should not be used at all since for typical AGN monitoring situations it gives errors that are twice as large as the GS method.

\item Methods of determining the lag by combining the assumption of a particular form of the response function or a model of the BLR with the assumption of a damped random walk model for the continuum variability, and then estimating the errors in the lag by a Markov chain Monte Carlo (MCMC) method (i.e., the JAVELIN method \citealt{Zu+11} and the method of \citealt{Li+13}) must not be used blindly and, in particular, it needs to be recognized that in a large fraction of cases the actual errors can be {\em substantially} larger than those given by the MCMC.
    
\item For planning observations, the expected accuracy in the lag determination can be estimated using Eqns. 1 and 2, with the caveat that observing campaigns shorter than a few major continuum events can bias lags to small values and do not allow for intrinsic changes in lags.

\end{enumerate}

Despite the problems with the current versions of some of the more recently proposed methods it is still a good idea to use more than one method as a check in potentially problematic cases. An example is where seasonal gaps are such that they cause ambiguity in lag determinations -- see, for example, the discussion of some individual objects in \citet{Zu+11}.

Investigation of schemes to determine lags needs to continue, but it is important to do as many independent checks on the accuracy of methods.  Experience to date suggest that gains in accuracy over the \citet{Gaskell+Sparke86} method will be modest at best.

\section*{ACKNOWLEDGEMENTS}

I am grateful to Ying Zu, Brad Peterson, Linda Sparke, Victor Oknyansky and Ski Antonucci for useful discussions.  I am also grateful to the anonymous referee for a careful reading which improved the paper.  Early stages of this research were supported by the GEMINI-CONICYT Fund of Chile through project N$^{\circ}$32070017 and through FONDECYT grant N$^{\circ}$1120957.

\bibliographystyle{aa}
\bibliography{Lag_errors_12}

\end{document}